\documentclass[aps,pre,groupedaddress,twocolumn,showpacs]{revtex4-1}

\usepackage{amsmath}
\usepackage{amsfonts}
\usepackage{amssymb}

\usepackage{color}
\usepackage{graphicx}
\usepackage{subfigure}

\renewcommand{\vec}[1]{{\mathbf #1}}
\newcommand{\evec}[1]{{\hat{\mathbf #1}}}

\begin{document}


\title{Effects of a temperature dependent viscosity on
 thermal convection in binary mixtures}


\author{Markus Hilt, Martin Gl\"assl, Walter Zimmermann}
\email[]{walter.zimmermann@uni-bayreuth.de}
\affiliation{Theoretische Physik I, Universit\"at Bayreuth, 95440 Bayreuth, GERMANY}


\date{\today}

\begin{abstract}
We investigate the effect of a temperature dependent viscosity on the onset
of thermal convection in a horizontal layer of a binary fluid mixture that is
heated from below. For an exponential temperature dependence of the viscosity, 
we find in binary mixtures as a function of a positive
separation ratio $\psi$ and
beyond a certain viscosity contrast a discontinuous transition between
two stationary convection modes having a different wavelength.
In the range of negative values of the separation ratio $\psi$, 
a (continuous or discontinuous) transition from an oscillatory to
a stationary onset of convection occurs beyond a certain viscosity contrast, 
and for large values of the viscosity ratio, the
oscillatory onset of convection is suppressed.
\end{abstract}


\maketitle

\section{Introduction}

Thermal convection occurs in fluids or gases heated from below
and it is a well known,  ubiquitous phenomenon \cite{Lappa:2010,Ball:98}.
It drives many important processes in geoscience
\cite{Turcotte:2002,Bercovici:2009,Busse:89.2,Davaille:2009.1}
or in the atmosphere  \cite{Houze:1994,StevensB:2005.1}, and it
is a central model system of nonlinear science  \cite{CrossHo,Cross:2009}.
Quite often, thermal convection can be described theoretically 
in terms of the so-called Oberbeck-Boussinesq (OB) approximation for 
a single component fluid, where constant material parameters are assumed,
except of the temperature-dependent density within the buoyancy term,
which is the essential driving force of convection. However, 
in nature, the viscosity may strongly depend on the temperature 
implying that models beyond the OB approximation have to be used
or convection takes place in fluid mixtures. Both degrees of 
freedom considerably affect convection, in particular near its onset. 
This work discusses the combination of both effects.

For a sufficiently large viscosity contrast between  
the lower warmer and the upper colder region of the
convection cell, related non-Boussinesq effects have be taken into account,
for instance, to model convection phenomena in the Earth's mantle
\cite{Palm:1960.1,JensenO:1963.1,Turcotte:1970.1,Booker:1978.1,Booker:1982.1,Busse:85.1,
White:1988.1,Richter:1983.1,Christensen:1991.1,Yuen:1995.1,Tackley:1996.1,Davaille:2009.1}.
First studies have shown, that a linear as well as a sinusoidal temperature dependence of the
viscosity of a fluid may lead to a  reduction of the onset of convection compared to
the case of a constant viscosity \cite{Palm:1960.1,JensenO:1963.1,Busse:85.1}.
In contrast, an exponential temperature dependence of the viscosity
can either lead to an enhancement or to a reduction
of the threshold  \cite{Booker:1982.1,White:1988.1},
depending on the strength of the  viscosity contrast.
Further, a spatially varying viscosity breaks the up-down symmetry in a convection layer
causing a subcritical convection onset to hexagonal patterns  \cite{White:1988.1}, 
and beyond the threshold, more complex convection regimes may be induced
in fluids having a temperature  dependent viscosity \cite{Davaille:2011.1}.

Research on convection in binary-fluid mixtures has a long tradition \cite{Turner:73,Huppert:1981.1}
with numerous applications in oceanography or geoscience  
\cite{Turner:73,Huppert:1981.1,Christensen:1993.1,Christensen:2010.1,HansenU:2012.1},
 nonlinear dynamics and bifurcations
\cite{CrossHo,Brand:84.1,Luecke:1992.1,Luecke:1998.1,Cross:88.2,Knobloch:88.1,Zimmermann:89.5,Zimmermann:93.1},
and more recently also to convection in colloidal suspensions
 \cite{Cerbino:2005.1,Pleiner:2008.1,Cerbino:2009.1,Glaessl:2010.1,Rehberg:2010.1,Glaessl:2011.1}.
In binary-fluid mixtures, the concentration field of one of the
two constituents enters the basic equations as an additional dynamic
quantity \cite{Legros:84,Landau6eng}: via the Soret effect ({\it thermophoresis}), 
a temperature gradient applied to a binary fluid mixture in a convection cell  
causes a spatial dependence of the concentration field, which couples
into the dynamical equation for the velocity field via the buoyancy term. 
The dynamics near the onset of convection in mixtures of alcohol and water
as well as in $^3$He/$^4$He mixtures is well investigated with a
good agreement between measurements and theory \cite{CrossHo,Luecke:1998.1}.
The possibility of a stationary as well as an oscillatory onset of 
convection in binary fluid-mixtures, including a so-called codimension-2 
bifurcation at the transition between both instabilities, caused additional attraction \cite{CrossHo}.

Although the temperature dependence of the viscosity as well as the two-component character
of fluids are considered to be of importance for modeling many phenomena in planetary science
\cite{Turcotte:2002,Bercovici:2009,Busse:89.2,Davaille:2009.1,Huppert:1981.1,Christensen:1993.1,Christensen:2010.1,HansenU:2012.1},  
the influence of a combination of both effects onto convection is still nearly unexplored \cite{Tanny:1995.1,Fleitout:2004.1}. 
As turbulent convection causes a homogenization of concentrations and of the temperature field
in the center of a convection cell, the impact of a combination of both effects is expected to be
less significant in the turbulent regime, but to be of particular importance
at the onset of convection, which is the focus of this work.

In Sec.~\ref{model}, we present the dynamical equations and  in Sec.~\ref{results_simple}, 
we reconsider the observation, that for a one-component fluid,
in the case of a linear temperature dependence of the viscosity
and a small viscosity contrast, one has a reduction of the onset
of convection, while there is an  enhancement of the threshold  for
an exponential temperature dependence. 
The influence of a temperature dependent viscosity on
the onset of convection in a binary mixture is considered in Secs.~\ref{results_binary1} and \ref{results_binary2},
both along the stationary branch as well as along the oscillatory branch,
including the codimension-2 point. Most striking, we find that the oscillatory
branch can be suppressed by strong viscosity contrasts.
In Sec.~\ref{conclusions}, the results are summarized and discussed.

\section{Basic equations and heat conducting state\label{model}}
Compared to the common basic equations for convection in binary fluid mixtures in Boussinesq approximation
\cite{Cross:88.2,Zimmermann:89.5}, we replace the constant viscosity  by a
 temperature dependent kinematic viscosity of a fluid $\nu=\nu_\infty\,\exp (\bar\gamma/T)$,
whereby we assume that both components of the mixture
have the same temperature dependence \cite{Raman:23,Eweell:37}. 
With the mean temperature in the convection cell,  $T_0$, and a 
Taylor expansion of the exponent around $T_0$ up to the leading order, the
viscosity takes the following form
\begin{align}
\label{nu}
\nu =\nu_0\,e^{-\gamma(T-T_0)} \, ,
\end{align}
where $\gamma=\bar\gamma/{T_0}^2$ and 
$\nu_0=\nu(T=T_0)=\nu_\infty\exp(\bar\gamma/{T_0})$. In a binary mixture, 
a  temperature dependent viscosity implies via $D\sim 1/\nu$
also a temperature dependent thermal diffusion constant $D$:
\begin{align}
\label{D}
D = D_0 \,e^{\gamma\, (T-T_0)} \, .
\end{align}
We would like to stress that this relation does not hold in general,
but is appropriate, when the dependence of the viscosity on the temperature
is roughly identical for both components or when  
the concentration of the second component is very small, such that the viscosity
of the mixture is almost exclusively determined by the first component. 
In Sec. III, we will restrict our analysis to these two cases.

The basic transport equations for an incompressible binary fluid mixture
involve a dynamical equation for the temperature field $T({\bf{r}},t)$,
the mass fraction of the second component $N({\bf{r}},t)$ and the fluid velocity
${\bf{v}}({\bf{r}},t)$:
\begin{subequations}
\label{basic}
\begin{align}
\nabla\cdot{\vec{v}} &= 0 \label{basic:1} \, ,\\
 \left(\partial_t + \vec{v}\cdot\nabla\right) T 
 &=  \chi \Delta T \, ,\\
  \left(\partial_t + \vec{v}\cdot\nabla\right) N
 &= \nabla \cdot \left(D \Big({\nabla} N +\frac{k_T}{{T_0}}{\nabla} {T} \Big)\right)\, ,\\
  \left({\partial}_t + {\vec{v}}\cdot{\nabla}\right) {\vec{v}} 
 &= -\frac{1}{\rho_0} {\nabla} {p} 
 + {\nabla} \cdot {\mathcal{S}}
 - \frac{\rho}{\rho_0} g \evec{e}_z \, .
\end{align}
\end{subequations}
Herein, 
\begin{align}
 \mathcal{S} 
&= \nu\left({\nabla} {\vec{v}} + ({\nabla} {\vec{v}})^T \right)
\end{align}
describes the stress tensor, $\chi$ denotes the thermal diffusivity of the mixture,
$k_T$ is the dimensionless thermal-diffusion ratio, that couples the
temperature gradient to the particle flux and is related to the Soret coefficient
$S_T$ via $k_T/T = N(1-N)S_T$, and $p({\bf{r}},t)$ denotes the pressure field.
As in the common Boussinesq approximation, we assume
that  $\chi$ and $k_T/T \sim N_0(1-N_0)S_T$ are constants and the dependence
of the density $\rho$ on $T$ and $N$ is taken into account only within the buoyancy
term, where we assume a linearized equation of state of the form 
\begin{align}
 \rho &= \rho_0 \left[1 - \alpha \left(T-T_0 \right) + \beta \left(N-N_0\right)\right] .
\end{align}
The Eqs.~(\ref{basic}) are completed by no-slip boundary conditions.
For a fluid that in the z-direction is confined between two impermeable,
parallel plates at a distance $d$ that are held at constant temperatures and 
extend infinitely in the x-y-plane, the following set of boundary conditions
results at $z=\pm d/2$:
\begin{subequations}
\begin{align}
T&=T_0 \mp \frac{1}{2}\,\delta T \, ,\\
0&=\partial_z N + \frac{k_T}{T_0}\,\partial_z T \, ,\\
0&=v_x=v_y=v_z=\partial_z v_z \, .
\end{align}
\end{subequations}
In the absence of convection (i.e., for ${\bf{v}}=0$), the time-independent
and with respect to the x-y-plane translational symmetric heat-conducting state is given by
\begin{subequations}
\label{conductive}
\begin{align}
T_{cond}(z)&=T_0-\delta T\,\frac{z}{d},\\
N_{cond}(z)&=N_0-\delta N\,\frac{z}{d}, \quad\text{with}\quad
                                 \delta N = -\frac{k_T}{T_0}\,\delta T . 
\end{align}
\end{subequations}
For the further analysis, it is convenient to separate this basic heat
conducting state from convective contributions setting
$T(\vec{r},t)=T_{cond}(z)+T_1(\vec{r},t)$ and
$N(\vec{r},t)=N_{cond}(z)+N_1(\vec{r},t)$. Making use of the rotational
symmetry in the fluid layer, we can restrict our analysis to
the x-z-plane and introduce a scalar velocity potential $F(x,z,t)$ via
\begin{align}
 v_x = -\partial_x^2 F,\qquad 
 v_z =  \partial_z \partial_x F,
\end{align}
with the help of which Eq.~\eqref{basic:1} is fulfilled by construction.
Rescaling distances by $d$, times by the vertical diffusion time $d^2/\chi$,
the temperature field $T$ by $\chi\nu_0 / \alpha g d^3$, the concentration
field $N$ by  $-k_T\chi\nu_0 / T_0 \alpha g d^3$ and the velocity potential
$F$ by $\chi d$, all material and geometry parameters are regrouped in
5 dimensionless parameters: the {\it Rayleigh number} $R$,
the {\it Prandtl number} $P$, the {\it Lewis number} $L$, and the 
{\it separation ratio} $\Psi$
\begin{align}
 P=\frac{\nu_0}{\xi},\,
 L=\frac{D_0}{\xi},\,
 R=\frac{\alpha g d^3}{\chi\nu_0}\,\delta T,\,
 \Psi=\frac{\beta k_T}{\alpha T_0}
\end{align}
are well known from common molecular binary-fluid mixtures and the fifth dimensionless
quantity
\begin{align}
 \Gamma = \frac{\chi\nu_0}{\alpha g d^3}\,\gamma
\end{align}
characterizes the viscosity contrast $\bar\nu$ between the viscosity at the upper
and the lower boundary via
\begin{align}
\label{eq_contrast}
 \bar\nu=\frac{\nu(z=+1/2)}{\nu(z=-1/2)}=e^{\Gamma R}.
\end{align}
In the following, we will discuss our results mainly in dependence on $\Psi$ and
$\Gamma$, whereas $P$ and $L$ are fixed to $P = 10$ and $L=0.01$, respectively. 

Finally, by introducing a rescaled temperature deviation $\theta=(R/\delta T)\,T_1$,
a rescaled concentration deviation $\tilde N_1=-(T_0 R/k_T \delta T)N_1$ as well as a 
rescaled velocity potential $f=1/(\chi d)F$ and using the combined function 
$\tilde c=\tilde N_1 -\theta$ instead of $\tilde N_1$, we obtain
\begin{subequations}
\label{eq_final}
\begin{align}
&\left( \partial_t - \Delta \right) \theta +R \partial_x^2 f
\nonumber \\
&\qquad= -\left(\partial_z\partial_x f \partial_x- \partial_x^2 f \partial_z \right)\theta\,,\\
%
&\partial_t c -L \nabla\cdot \left(e^{\Gamma (-Rz+\theta)}\nabla c\right)+\Delta \theta 
\nonumber \\
&\qquad= -\left(\partial_z\partial_x f \partial_x- \partial_x^2 f \partial_z \right)c \,,\\
&\partial_t \Delta\partial_x f - P\Delta\left(e^{-\Gamma (-Rz+\theta)}\Delta\partial_x f\right) 
\nonumber \\
&\, +P\Psi\partial_x c+P(1+\Psi)\partial_x\theta
\nonumber \\
&\,+2P\left[\left(\partial_z^2 e^{-\Gamma (-Rz+\theta)}\right)\partial_x^2 +\left(\partial_x^2 e^{-\Gamma (-Rz+\theta)}\right)\partial_z^2 \right]\partial_x f
\nonumber \\
&\,-4P \left(\partial_x\partial_z e^{-\Gamma (-Rz+\theta)}\right)\partial_x^2\partial_z f
\nonumber\\
&\qquad= -\left(\partial_z\partial_x f \partial_x- \partial_x^2 f \partial_z \right)\partial_x f \,,
\end{align}
\end{subequations}
together with the no-slip, impermeable boundary conditions
\begin{align}
\label{boundarycond}
 \theta = \partial_z c = \partial_x f = \partial_z\partial_x f = 0 \quad\text{at}\quad z=\pm 1/2,
\end{align}
where, for simplicity, all tildes have been suppressed.

\section{Onset of convection\label{results}}

The parameters at the onset of convection are determined by a linear stability analysis of the basic, nonconvective
state $\theta=c=f=0$, as for instance described in more detail in Ref.~\cite{Glaessl:2010.1}. 

For this purpose, the linearized equations 
%
\begin{subequations}
\label{eq_linear}
\begin{align}
\partial_t \theta 
&= \Delta \theta - R \partial_x^2 f\\
\partial_t c 
&= -\Delta \theta +L \nabla\cdot \left(e^{-\Gamma Rz}\nabla c\right)
 \\
\frac{1}{P}\partial_t \Delta\partial_x f 
&= \, -\Psi\partial_x c - (1+\Psi)\partial_x\theta \nonumber \\
& \qquad + \Delta\left(e^{\Gamma Rz}\Delta\partial_x f\right) 
  - 2 \Gamma^2 R^2 e^{\Gamma Rz}\partial_x^2 \partial_x f 
\end{align}
\end{subequations}
are solved by a Fourier ansatz
along the horizontal direction:
$\left(\theta,c,f \right)=\left(\bar{\theta}(z),\bar{c}(z),\bar{f}(z)\right)\,\exp(i\,k\,x+\sigma\,t)$.
The $z$-dependence of the fields $\bar{\theta}(z)$, $\bar{c}(z)$, $\bar{f}(z)$ 
are expanded with respect to orthogonal polynomials that fulfill the
boundary conditions in Eq.~(\ref{boundarycond}). By a projection of the linear
equations onto these polynomials (Galerkin-Method, see, e.g., Refs. \cite{Busse:1974.2,Canuto:1987.1,Pesch:1996.1}), 
the dynamical equations are transformed into an eigenvalue problem. 
%
By the condition  $\mathrm{Re}(\sigma)=0$, the neutral curve $R_0(k)$ for the Rayleigh number is determined,  whose 
minimum $(R_c,k_c)$ at the critical Rayleigh number $R_c$ and the critical wavenumber $k_c$ determines the onset 
of convection. With $\omega_c=\mathrm{Im}(\sigma)$, we denote the frequency at the threshold of the oscillatory onset 
of convection.

\subsection{Simple fluids ($\psi=0$)\label{results_simple}}

At first, let us concentrate on the effect of an exponentially temperature-dependent
viscosity on the onset of convection for one-component fluids ($\psi=0$). 

\begin{figure}[ttt]
 \includegraphics[width=0.95\columnwidth]{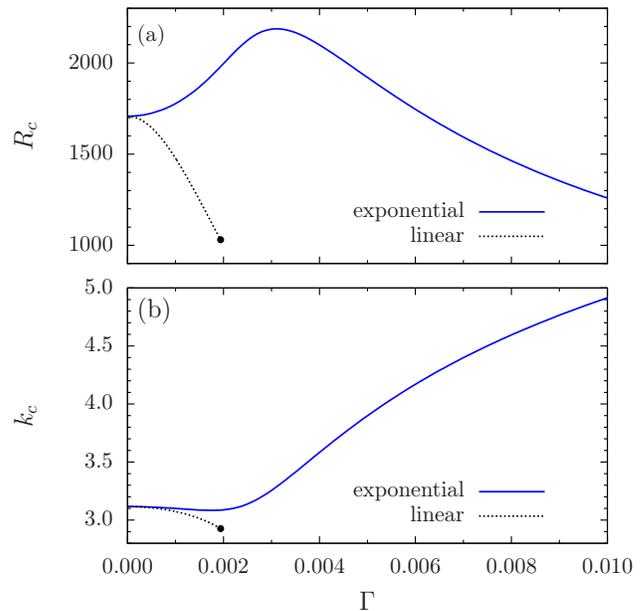}
\caption{\label{fig_simple1}
(color online) (a) The critical Rayleigh number $R_c$ and (b) the critical
wavenumber $k_c$ for a one-component fluid as a function of $\Gamma$. 
The solid line marks an exponential temperature-dependence of the
viscosity, the dotted line represents a linear one. The dotted line
ends at $\Gamma\approx 1.941\,\times\,10^{-3}$ (black point),
where the viscosity becomes negative at the upper boundary.}
\end{figure}

\begin{figure}[ttt]
\includegraphics[width=0.98\columnwidth]{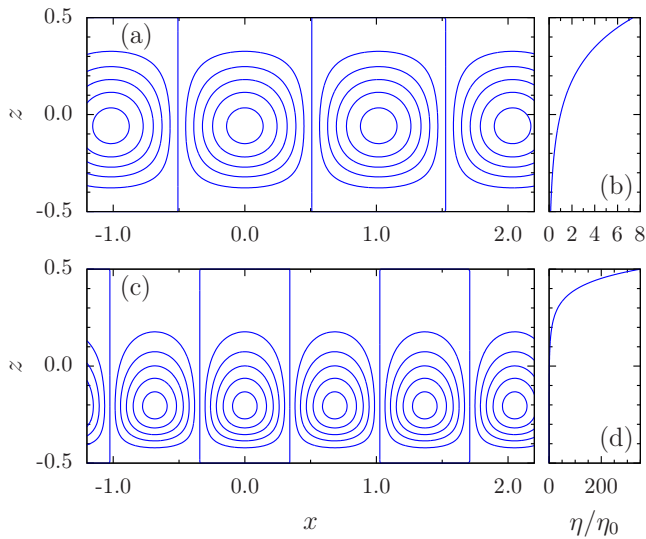}
\caption{\label{fig_simple2}
(color online) Contour lines of the velocity potential $f$ 
at the onset of convection for  (a) $\Gamma=0.002$   
and (c)  $\Gamma=0.008$, respectively.
Parts (b) and (d) show the corresponding spatial dependence of the
viscosity $\eta(z)/\eta_0$. The critical values are $k_c=3.09$ and
$R_c=1997$ in (a, b) and  $k_c=4.59$ and  $R_c=1464$ in (c, d). 
}
\end{figure}

For this case, the critical Rayleigh number $R_c$ and the corresponding critical 
wavenumber $k_c$ are shown in Fig.~\ref{fig_simple1} as a function of $\Gamma$ (solid lines).
Both quantities reveal a non-monotonic dependence on $\Gamma$, similar to the results
reported in Refs.~\cite{Booker:1982.1,Kameyama:2013.1}:
while for small $\Gamma$, $R_c$ rises compared to the case of a constant viscosity,
the threshold is reduced in the limit of large $\Gamma$. 
This contrasts to related studies \cite{Busse:85.1}, where a linear temperature-dependence
of the viscosity has been assumed and which predict a 
monotonic decrease of the threshold with rising viscosity contrast.
However, we can reproduce that result by a linear approximation of the exponential
terms in Eqs.~\eqref{eq_final}, which is also shown in Fig.~\ref{fig_simple1} (dotted lines)
and which clearly demonstrates the importance of terms higher than the leading linear order.
The velocity potential $f$ at the onset of convection is shown in Fig. \ref{fig_simple2} for two values of $\Gamma$.
The stronger the viscosity varies in space, the more the center of the convection rolls is shifted towards 
the lower boundary and the more the fluid motion is suppressed near the upper boundary, where a highly
viscous layer forms.


\subsection{Binary mixtures with positive Soret effect ($\psi>0$)}\label{results_binary1}


In the range of a positive Soret effect, i.~e. $\psi>0$, convection
in binary fluid mixtures sets in stationary for all $\Gamma$, just
as  in the case of a constant viscosity 
\cite{Cross:88.2,Knobloch:88.1,Zimmermann:89.5,Zimmermann:93.1}. 
\begin{figure}[ttt]
\includegraphics[width=0.98\columnwidth]{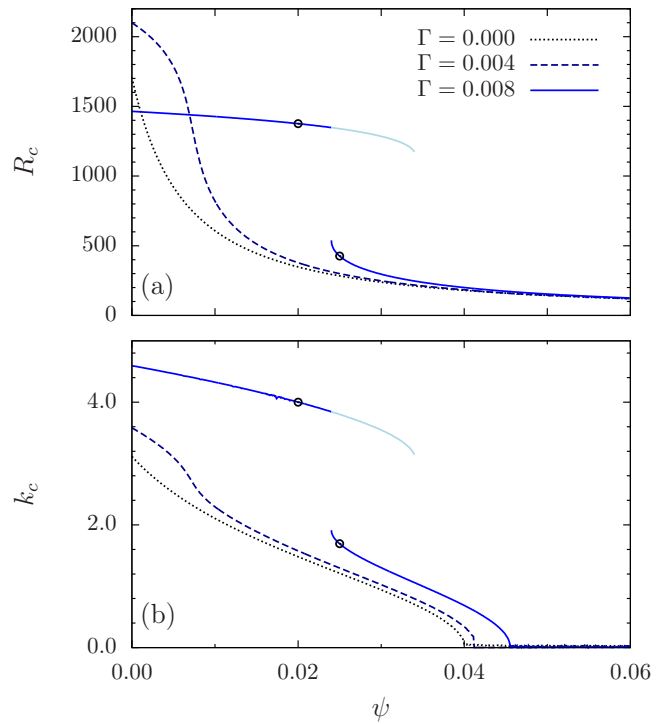}
\caption{\label{fig_pos1}
   (color online) Critical values (a) $R_c(\psi)$ and (b) $k_c(\psi)$
   for $\Gamma=0.000$, $0.004$, and $0.008$. 
   The circles mark those values of $\psi$, for which neutral
   curves are shown in Fig.~\ref{fig_pos2}.
}
\end{figure}

Fig.~\ref{fig_pos1} shows $R_c$ and $k_c$ as functions of $\psi$ for two
representative finite values of $\Gamma$ as well as for the limiting case $\Gamma=0$. 
For moderate values of $\Gamma$ (dashed lines), $R_c$ and $k_c$ are higher than for $\Gamma=0$
(dotted lines) and their behavior as functions of $\psi$, in particular the shift of $k_c$ towards zero
for rising values of $\psi$, is pretty similar to that for $\Gamma=0$.
However, for higher values of $\Gamma$ (solid lines) and small $\psi$, the threshold is reduced
compared to $\Gamma=0$, which is similar to the case of a simple fluid as shown in Fig.~\ref{fig_simple1}. 
In addition, for large $\Gamma$, the decay of $R_c$ as a function of $\psi$ becomes much weaker and, most importantly, 
at a certain value of $\psi$, the threshold discontinuously jumps down to much lower values,
which are comparable to those for $\Gamma=0$. 

\begin{figure}[ttt]
\includegraphics[width=0.98\columnwidth]{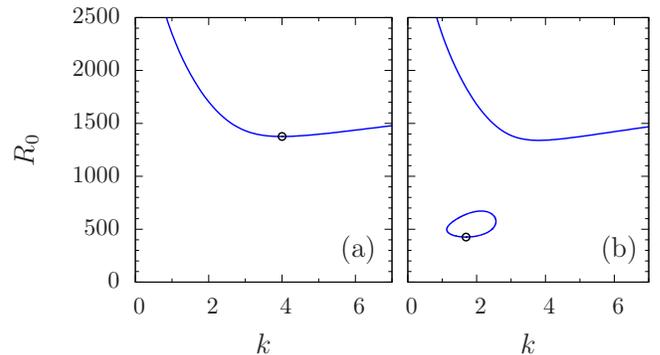}
\caption{\label{fig_pos2}
   (color online) Neutral curves corresponding to the circles in Fig. \ref{fig_pos1}
   with (a) $\psi=0.02$ and (b) $\psi=0.025$ for $\Gamma=0.008$.
   The points mark the minima of the neutral curves. 
%
}
\end{figure}

To understand this discontinuous behavior, Fig.~\ref{fig_pos2} shows the neutral
curves $R_0(k)$ for two values of $\psi$, which are to the right or left of the jump,
respectively, and which are marked by circles in Fig.~\ref{fig_pos1}. 
For the larger value of $\psi$ [cf. Fig.~\ref{fig_pos2}(b)], an additional region
of stationary instability forms in the $(R,k)$-plane with a minimum at lower Rayleigh
numbers, which explains the discontinuity shown in Fig.~\ref{fig_pos1}.

\begin{figure}[ttt]
\includegraphics[width=0.98\columnwidth]{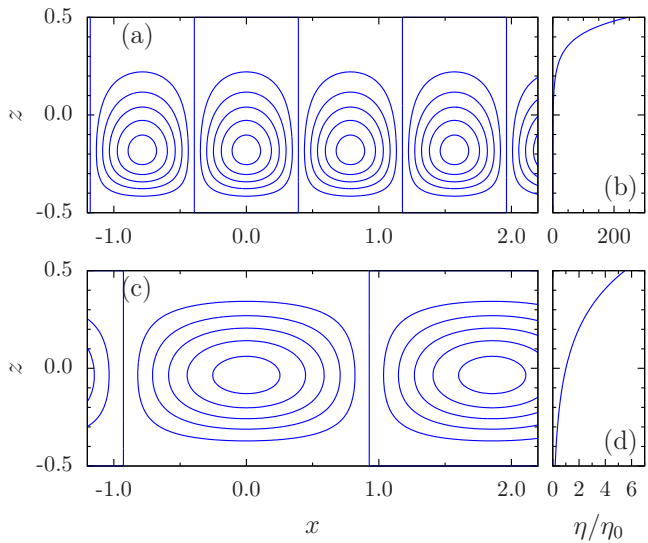}
\caption{\label{fig_pos3}
    (color online) Contour lines of the velocity potential $f$
     at the onset of convection corresponding to 
    (a) Fig. \ref{fig_pos2}(a) ($\psi=0.020$, $\Gamma=0.008$, $k_c \cong 4.00$, $R_c \cong 1376$) and 
    (c) Fig. \ref{fig_pos2}(b) ($\psi=0.025$, $\Gamma=0.008$, $k_c \cong 1.69$, $R_c \cong 426$).
    (b) and (d) depict the corresponding decay of the viscosity.
}
\end{figure}

As the viscosity contrast [cf. Eq.~\eqref{eq_contrast}] at the onset of convection is given by the 
product of $\Gamma$ and $R_c$, the jump in the critical Rayleigh number leads for $\psi$ close to
the discontinuity to a strong change in the viscosity contrast at the threshold. 
This finally leads to very different velocity fields at the onset of convection for values of $\psi$ that are 
to the right or to the left of the jump, which is illustrated by the velocity potential in Figs.~\ref{fig_pos3} (a)
and~(c), respectively. For the smaller value of $\psi$, $R_c$ is higher [cf. Fig.~\ref{fig_pos2}(a)], 
leading to a stronger viscosity contrast [cf. Fig.~\ref{fig_pos3}(b)] and therefore to a pronounced shift
of the flow field towards the lower boundary [cf. Fig.~\ref{fig_pos3}(a)]. In contrast, for the larger
value of $\psi$, the threshold $R_c$ is smaller [cf. Fig.~\ref{fig_pos2}(b)], the viscosity contrast
is much weaker [cf. Fig.~\ref{fig_pos3}(d)], and thus, there is only a slight shift of the convection
rolls [cf. Fig.~\ref{fig_pos3}(c)]. Further, the different lateral extension of the role structure
shown in Figs.~\ref{fig_pos3}(a) and~(c) reflects the jump in $k_c$ [cf. Fig.~\ref{fig_pos1}(b)].  

\subsection{Binary mixtures with negative Soret effect ($\psi<0$)}\label{results_binary2}

The most interesting effect of a strongly temperature-dependent viscosity occurs in
the range of a negative Soret effect, i.~e., for $\psi<0$, 
where with increasing values of $\Gamma$, the divergence of the stationary instability (in the case of
a constant viscosity \cite{Cross:88.2,Knobloch:88.1,Zimmermann:89.5,Zimmermann:93.1})
vanishes. Further, beyond a certain L-dependent value of $\Gamma$,
the onset of convection is no longer oscillatory for all $\psi<0$, as it is known in the case of a constant viscosity.
Instead, at strongly negative values of $\psi$, the oscillatory instability is replaced by a stationary one.
Depending on the strength of the exponential temperature-dependence of the viscosity, the transition from
a Hopf-bifurcation to a stationary instability with decreasing $\psi$ can show a discontinous or a continous
threshold behavior.

\subsubsection{Discontinous transition from an oscillatory to a stationary instability}

\begin{figure}[ttt]
\includegraphics[width=0.98\columnwidth]{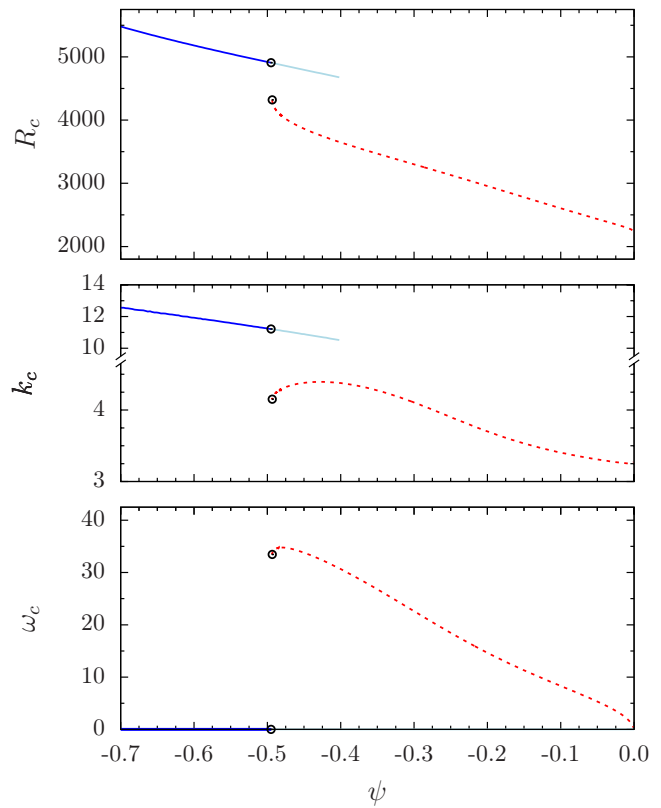}
\caption{\label{fig_neg_a1} 
    (color online) (a) Critical Rayleigh number $R_c$,
    (b) critical wavenumber $k_c$, and 
    (c) critical frequency $\omega_c$
    as functions of $\psi < 0$ for $\Gamma=0.003$.
}
\end{figure}
\begin{figure}[hhh]
\includegraphics[width=0.98\columnwidth]{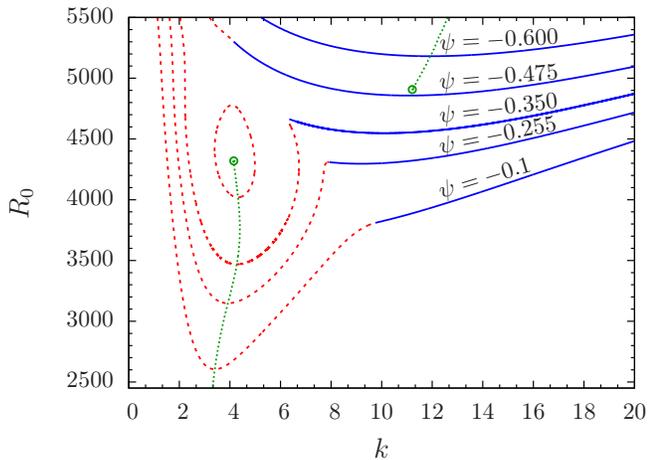}
\caption{
\label{fig_neg_a2} 
   (color online) Neutral curves for $\Gamma=0.003$ and different $\psi<0$ as indicated.
   The green dotted line marks the position of the absolute minima of the neutral curves. 
   The black line denotes the critical value $\Psi_c \cong -0.5$, where the transition
   from an oscillatory to a stationary instability takes place. At that point, the minimum
   of the neutral curves shows a discontinous jump in $k_c$ and $R_c$ (points). 
}
\end{figure}

\begin{figure}[hhh]
\includegraphics[width=0.98\columnwidth]{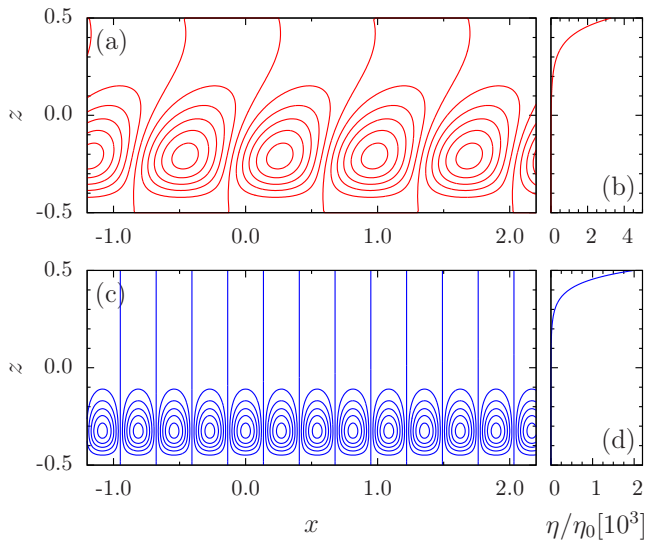}
\caption{\label{fig_neg_a3} 
   (color online) Contour lines of the velocity potential $f$ at the onset of convection 
   corresponding to Fig. \ref{fig_neg_a2} for 
   (a) $\psi=-0.45$, $k_c \cong 4.38$, $R_c \cong 3864$ and 
   (c) $\psi=-0.55$, $k_c \cong 11.63$, $R_c \cong 5060$.
   (b) and (d) depict the corresponding decay of the viscosity.
}
\end{figure}

For moderate values of $\Gamma$, the transition between both types of instabilities is characterized by a
discontinous jump in $R_c$, $k_c$, and $\omega_c$, as exemplarily illustrated in Fig.~\ref{fig_neg_a1} for $\Gamma=0.003$.
The corresponding neutral curves $R_0(k)$ for different $\psi$ are shown in Fig.~\ref{fig_neg_a2}, where 
red lines indicate those parts of the neutral curves where the frequency $\omega$ is finite, while blue lines represent a stationary 
instability with $\omega=0$. For small $|\psi|$, the minima of the neutral curves (green line) 
belong to an oscillatory instability. However, with increasing $|\psi|$, this region transforms into
an oscillatory island, which finally disappears, while the stationary branch of the curve, which shows a minimum at 
larger values of $k$, remains. In consequence, for even larger $|\psi|$, convection sets in stationary 
at a higher threshold and a considerably increased critical wavenumber. These changes are directly reflected 
in the velocity potential at the onset of convection, as shown in Fig.~\ref{fig_neg_a3} (a,c): while Fig.~\ref{fig_neg_a3}(a)
shows travelling waves in the regime of the oscillatory instability, Fig.~\ref{fig_neg_a3}(c) displays stationary
convection rolls with a much smaller lateral width (due to the jump in $k_c$), which are also much more shifted
to the lower boundary [due the higher threshold and, hence, the more pronounced viscosity contrast, cf.
Figs.~\ref{fig_neg_a3}(b,d)]. 

\subsubsection{Continous transition from an oscillatory to a stationary instability}

\begin{figure}[ttt]
\includegraphics[width=0.98\columnwidth]{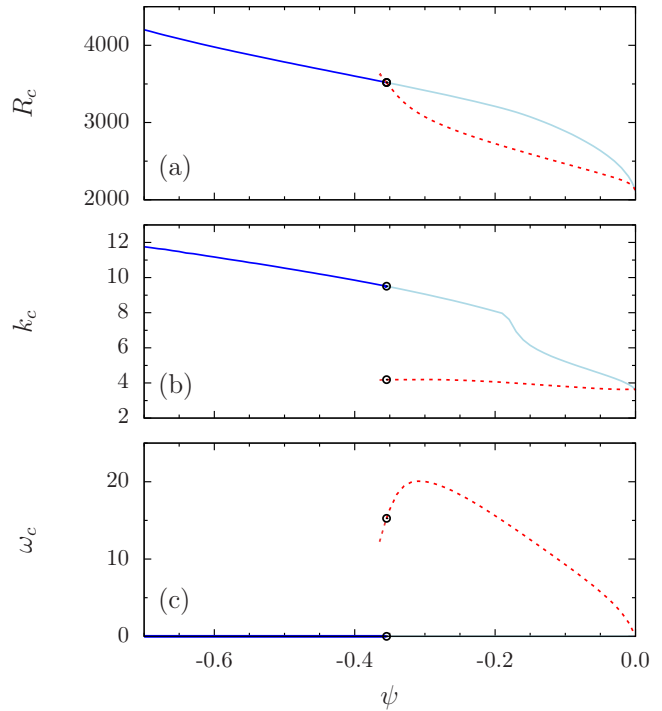}
\caption{\label{fig_neg_b1} 
    (color online) (a) Critical Rayleigh number $R_c$,
    (b) critical wavenumber $k_c$, and
    (c) critical frequency $\omega_c$
    as functions of $\psi<0$ for $\Gamma=0.004$.
}
\end{figure}

\begin{figure}[hhh]
\includegraphics[width=0.98\columnwidth]{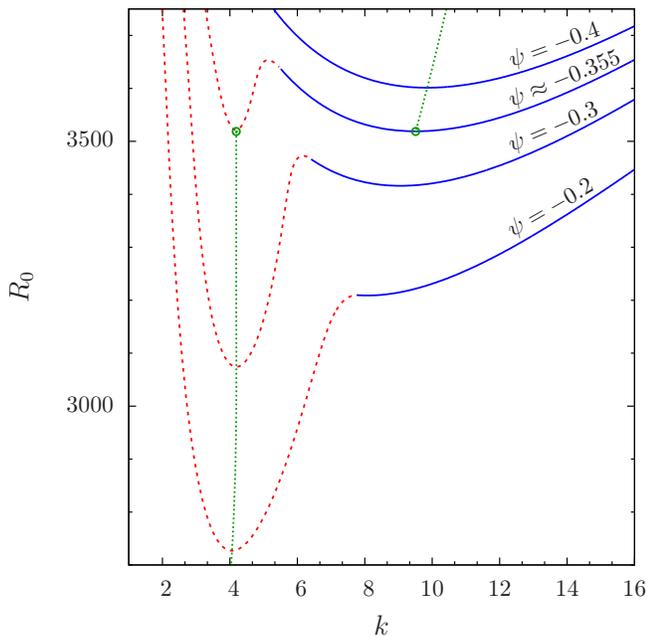}
\caption{
  \label{fig_neg_b2} 
   (color online) Neutral curves for $\Gamma=0.004$ and different $\psi<0$ as indicated.
   The green dotted line marks the position of the absolute minima of the neutral curves, 
   the thick line denotes the critical value of $\psi_c \cong -0.355$, where
   the transition from an oscillatory to a stationary instability takes place. 
   At that point, the minimum
   of the neutral curves shows a discontinous jump in $k_c$ (points). 
}
\end{figure}

For larger values of $\Gamma$, the transition between the Hopf and the stationary bifurcation 
is still characterized by jumps in the critical wavenumber and the critical frequency, but
does no longer show a discontinuity in the threshold, as illustrated in Fig.~\ref{fig_neg_b1}
for $\Gamma = 0.004$. 
Instead of forming an oscillatory island, that, for rising $|\psi|$, is eventually disappearing, here, as displayed
in Fig.~\ref{fig_neg_b2}, for rising
$|\psi|$, the minimum of the oscillatory branch of the neutral curves moves higher and higher. 
At a certain value of $\psi$, the minima of the oscillatory and stationary branches are of
equal height and with further increasing $|\psi|$, the minimum of the stationary branch is
finally lower and determines the onset of convection. The changes of the velocity potential
near this new codimension-2-point are similar to those depicted in Fig.~\ref{fig_neg_a3}. 
The stationary branch of the critical Rayleigh number, shown in Fig~\ref{fig_neg_b1}(a) (blue line)
in the range of $\psi<0$ is continued by the corresponding curve ($\Gamma=0.004$) in Fig~\ref{fig_pos1}(a)
to the range $\psi>0$.

A further interesting difference between the scenarios shown in Figs.~\ref{fig_neg_a1}-\ref{fig_neg_a3}
and Figs.~\ref{fig_neg_b1}-\ref{fig_neg_b2} is that for increasing $\Gamma$, the change from oscillatory to stationary convection
takes place at a smaller value of $|\psi|$. When further increasing $\Gamma$, this trend 
continues, i.e., for rising strength of the exponential temperature dependence of the
viscosity, the region in the parameter range $\psi<0$, where convection sets in via
a Hopf bifurction, becomes smaller and smaller. 

\section{Summary and Conclusions \label{conclusions}}
The parameters at the onset of convection are determined
 in a binary fluid mixture where the viscosity depends exponentially
on the temperature. 

As explicitely shown for a single component fluid, the critical values at the onset of
convection behave as a function of the viscosity difference between the lower, warmer and
the upper, coulder boundary differently for a linear temperature dependent and
an exponentially temperature dependent viscosity, respectively. 

In the range of a positive separation ratio $\psi$, we find, as a function of $\psi$,
for larger values of the viscosity contrast  a  discontinuous change of the critical 
Rayleigh number as well as of the critical wavelength of the convection rolls, in
contrast to their continuous behavior in the range of a constant
viscosity and small values of the viscosity contrast.

The strongest qualitative influence of an exponentially dependent 
viscosity at the onset of convection we find in the range of negative values of the
separation ratio $\psi$. In molecular binary mixtures, for $\psi<0$, below the 
onset of concevtion, the minor and heavier component of the fluid mixture is, via 
the Soret effect, enriched near the lower and warmer boundary. 
In geophysical applications, where also double diffusive models are applied,
the Soret effect does not play a very strong role, but due to 
gravitation, the heavier minor component of the mixture is similarly accumulated 
in the lower warmer range of the convection layer.
For molecular binary fluids, such as water-alcohol mixtures,
it is common that in closed convection cells, one
has an oscillatory onset of convection in the range $\psi<0$.
However, beyond the threshold, the concentration gradient is
quickly reduced by the convective motion, which soon 
leads to a stationary convection pattern again \cite{Luecke:1998.1}.
In the case of an exponentially temperature dependent viscosity
of the binary mixture, we find in the range $\psi<0$ the surprising effect, 
that with increasing values of the viscosity  contrast, already the onset
of convection changes from an oscillatory to a stationary one and that 
the range $\psi<0$, in which the onset of convection
is still oscillatory, shrinks with increasing viscosity contrasts.  
According to this result for closed convetion cells, we expect also
in model systems, where one has nonvanishing currents of the minor component
through the lower boundary  \cite{Fleitout:2004.1,Christensen:2010.1} and
that are of importance for geohysical situations,
a stationary onset of convection.

\section*{Acknowledgment}

We are grateful to Georg Freund for instructive discussions about how to implement
the Galerkin method in an efficient way.


\end{document}